\documentclass[preprint,aps]{revtex4}
\usepackage{epsfig}

\def\rhovec{\mbox{\boldmath $\rho$}}

\newcommand{\beq}{\begin{eqnarray}}
\newcommand{\eeq}{\end{eqnarray}}

\begin{document}

\title{Four-body structure of neutron-rich
hypernucleus $^6_{\Lambda}$H}

\author{E.\ Hiyama$^1$, S. Ohnishi$^{1,2}$, M. Kamimura$^3$, 
and Y. Yamamoto$^1$}

\address{$^1$Nishina Center for
Accelerator-Based Science,
Institute for Physical and Chemical
Research (RIKEN), Wako
351-0198, Japan}

\address{$^2$Department of Physics, Tokyo Institute of Technology,
Meguro 152-8551, Japan}

\address{$^3$Department of Physics, Kyushu University,
Fukuoka 812-8581, Japan}

%

\begin{abstract}
The structure of heavy hyperhydrogen $^6_{\Lambda}$H
is studied within the framework of a $tnn\Lambda$ four-body cluster model.
Interactions among the constituent subunits are
determined so as to reproduce reasonably 
well the observed low-energy properties
of the $tn, t\Lambda$ and  $tnn$ subsystems.
As long as we reproduce the
energy and width of $^5$H within the error bar,
the ground state of $^6_{\Lambda}$H is obtained as
a resonant state.
\end{abstract}

\maketitle

\parindent 20 pt

\section{Introduction}
One of the research goals in hypernuclear physics
is to study new dynamical features by 
injecting a $\Lambda$ particle into a nucleus.
For example, since there is no Pauli principle
between nucleons and a $\Lambda$ particle,
the $\Lambda$ participation gives rise
to more bound states and significant contraction 
of nuclear cores, especially in light systems.
Such a dynamical change in light hypernuclei
has been studied mostly in systems
composed of a stable nucleus and an attached 
$\Lambda$ particle \cite{Motoba83,Bando90}.
In light nuclei near the neutron drip line,
there have been observed 
interesting phenomena concerning neutron halos.
A corresponding subject in hypernuclear physics is to 
focus on structures of neutron-rich $\Lambda$ hypernuclei.
If a $\Lambda$ particle is added to such nuclei
with weakly-bound neutrons or resonant ones,
a resultant hypernucleus will become more stable against
neutron decay.
Thanks to this gluelike role
of an attached $\Lambda$ particle,
there is a new chance to produce a hypernuclear 
neutron-(proton-) state, if the core nucleus has 
an unbound (resonant) nucleon state with an appropriate 
energy above the nucleon-decay threshold.
Another interest is to extract information about
$\Lambda N -\Sigma N$ coupling effects in hypernuclei.
It is thought that this effect might play an important role
in neutron-rich $\Lambda$ hypernuclei, because the
total isospin becomes large.

For such an aim,
the structure of $^6_{\Lambda}$He,
$^7_{\Lambda}$He, $^7_{\Lambda}$Li and $^7_{\Lambda}$Be
was investigated \cite{Hiyama96, Hiyama09}
using an $\alpha$ cluster model, and it was pointed out that
these hypernuclei were of neutron- or proton-halo structure.
One of them, the neutron-rich $\Lambda$ hypernucleus
$^7_{\Lambda}$He, was observed in the $(e, e'K^+)$ reaction at Jlab 
\cite{Nakamura2013}, and an observed
$\Lambda$ separation energy of $B_{\Lambda}=5.68 \pm 0.03 \pm 0.25$ MeV
was reported.

On the other hand, to produce a neutron-rich hypernucleus $^{10}_{\Lambda}$Li,
the double-charge exchange $(\pi^-, K^+)$ reaction on a
$^{10}$B target  was performed at KEK~\cite{Saha05}.
On the basis of the result of this experiment,
Umeya and Harada~\cite{Umeya2011} calculated the structure of the Li-isotope 
$\Lambda$ hypernuclei
within a shell-model framework, and they stressed that
$\Lambda N -\Sigma N$ coupling effects play a moderate role for the 
$\Lambda$-binding energy as the number of neutrons 
increases.

Recently, a FUNUDA experiment~\cite{FINUDA,FINUDA-full} reported an 
epoch-making observation of superheavy hydrogen-$\Lambda$ 
hypernucleus $^6_{\Lambda}$H as a bound state with 
$B_{\Lambda}=4.0 \pm 1.1$ MeV.
$^6_{\Lambda}$H is a neutron-rich system including 
four neutrons and only a single proton, which goes far toward the 
neutron drip line. 
Others have previously analyzed the structure of this hypernucleus.
For example, Dalitz {\it et al.}
predicted that $^6_{\Lambda}$H was a bound state
based upon shell model arguments \cite{Dalitz}.
Akaishi {\it et al.} suggested that 
coherent $\Lambda N-\Sigma N$ coupling is important 
for the $\Lambda$ binding in $^6_{\Lambda}$H \cite{Akaishi}.

With motivation from these experimental and theoretical studies, 
a search (Experiment E-10) for $^6_{\Lambda}$H using the 
double-charge exchange $(\pi^-, K^+)$ reaction at J-PARC~\cite{Sakaguchi}
has been performed, and the analysis is now in progress.
Thus, it is quite timely to study the structure of $^6_{\Lambda}$H 
within a realistic approach.

It should be noted that the ground state of the core nucleus $^5$H was
observed as a resonant state with a broad width,
$E=1.7 \pm 0.3$ MeV ($\Gamma=1.9 \pm 0.4$ MeV)~\cite{H5}, with
respect to the $tnn$ three-body breakup threshold.
 We further note that
the $\Lambda$ separation energy, 
$B_\Lambda$, depends strongly on the spatial structure (size) of the core 
nucleus~\cite{Hiyama2000PRL,Hiyama2012PTP}.
Therefore, before calculating the binding energy of $^6_{\Lambda}$H,
it is essential to reproduce the observed data for $^5$H.
The resonant structure of $^5$H has been studied in the literature
within the framework of a $tnn$ three-body model;
it was found that $^5$H was described well 
by using such a model \cite{Zhukov2000,Diego}.
Therefore, it is reasonable to
employ a $tnn\Lambda$ four-body model for the
study of $^6_{\Lambda}$H in the present work.
In order to discuss the energy and width of 
$^5$H on the basis of the $tnn$ three-body model,
we employ the complex scaling method (CSM), 
a powerful tool for analyzing such three-body resonance 
states~\cite{CSM-ref1,CSM-ref2,CSM-ref3,Ho,Moiseyev}.

All the  interactions
among subunits (triton, two neutrons and $\Lambda$)
are chosen to reproduce the appropriate
low energy properties, such as
binding energies and scattering phase shifts for each of the
subsystems composed of two and
three subunits.
Using these interactions,
we calculate the energy of $^6_{\Lambda}$H
based on the $tnn\Lambda$ four-body model
and discuss whether or not 
$^6_{\Lambda}$H exists as a bound system.

In Sect.~2, the method used in the four-body calculation for
the $tnn\Lambda$ system is described.
In Sect.~3, we explain the interactions employed.
The calculated results and discussion are
presented in Sect.~4.
A summary is given in Sect.~5.

\section{Model}

\subsection{The $^6_{\Lambda}$H hypernucleus}

In this work the hypernucleus $^6_{\Lambda}$H
is considered 
as a triton cluster,
a $\Lambda$ particle, and two valence neutrons.
The triton cluster is considered to be an
inert core (an elementary particle) and to have a wave function $\Phi(t)$
with a $(0s)^3$ shell model configuration, .

Nine sets of Jacobian coordinates
for the four-body system of $^6_{\Lambda}$H
are illustrated in Fig.1; additionally,
we further take into account
the antisymmetrization
between the two neutrons.

\begin{figure}[htb]
\begin{center}
\epsfig{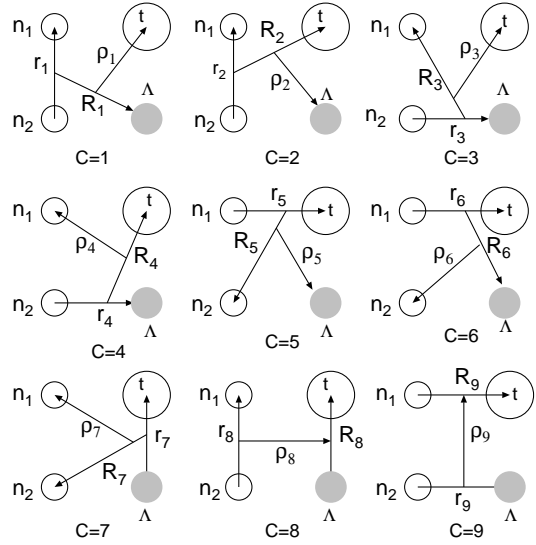}
\end{center}
\caption{Jacobi coordinates for all
the rearrangement channels ($c=1 \sim 9$)
of the $tnn\Lambda$ four-body system.
The two nucleons are to be antisymmetrized within each channel.}
\label{fig:h6l-jacobi}
\end{figure}
%
The total Hamiltonian
and the Schr\"{o}dinger equation  
are given by
\begin{eqnarray}
&& \qquad \qquad
( H - E ) \, \Psi_{JM,TT_z}(^6_{\Lambda}{\rm Z})  = 0 ,\\  
\label{eq:schr6}
&& \!\!\!\!\!\!\!\!\!\! \!\!\!\!\!\!  H=T+ \!V_{n_1 n_2}
      + \!\sum_{i=1}^2 ( V_{\Lambda n_i}   
     \! + \!V_{t n_i}) +V_{tn_1n_2}   
      + \!V_{t \Lambda} ,  
\label{eq:hamil6}
\end{eqnarray}
where $V_{t n_i}$ and $V_{n_1 n_2}$ are the interactions between
the triton and $i$-th neutron and the one between two neutrons, respectively.
$V_{t \Lambda}$ is the triton-$\Lambda$ interaction and
$V_{tn_1n_2}$ is the triton-$nn$ three-body force. 
These interactions  are explained in the next section.
The total wave function is described
as a sum of amplitudes for all the rearrangement channels
shown in 
Fig.~\ref{fig:h6l-jacobi} and $LS$ coupling scheme:
 \begin{eqnarray}
   \Psi_{JM}(^6_{\Lambda}{\rm Z}) 
 =  \sum_{c=1}^{7}\!\!\!\! &&\!\!\!\!
      \sum_{nl, NL, \nu \lambda}
       \sum_{IK} \sum_{ss'S}
       C^{(c)}_{nl,NL,\nu\lambda, IK, ss'S}\: \Phi(t) \nonumber  \\
      &\times&    {\cal A}
      \Big\{ \Big[
        \big[ \, [ \phi^{(c)}_{nl}({\bf r}_c)
         \psi^{(c)}_{NL}({\bf R}_c)]_I
      \,  \xi^{(c)}_{\nu\lambda} (\mbox{\boldmath $\rho$}_c)
        \big]_{K} \nonumber  \\
     &\times&    
      \big[ \big[ \,
     [ \chi_{\frac{1}{2}}(n_1)
       \chi_{\frac{1}{2}}(n_2)
       ]_s \chi_{\frac{1}{2}}(t) \big]_{s'} \chi_{\frac{1}{2}}(\Lambda) \big]_S 
           \Big]_{JM} \; \Big\},
\nonumber \\
 + \sum_{c=8}^{9}\!\!\!\!&&\!\!\!\!
       \sum_{nl, NL, \nu \lambda}
       \sum_{IK} \sum_{ss'S}
       C^{(c)}_{nl,NL,\nu\lambda, IK, ss'S}\: \Phi(t) \nonumber  \\
      &\times&    {\cal A}
      \Big\{ \Big[
        \big[ \, [ \phi^{(c)}_{nl}({\bf r}_c)
         \psi^{(c)}_{NL}({\bf R}_c)]_I
      \,  \xi^{(c)}_{\nu\lambda} (\mbox{\boldmath $\rho$}_c)
        \big]_{K} \nonumber  \\
     &\times&    
      \big[ 
     [ \chi_{\frac{1}{2}}(n_1)
       \chi_{\frac{1}{2}}(n_2)
   ]_s [ \chi_{\frac{1}{2}}(t) \chi_{\frac{1}{2}}(\Lambda) ]_{s'}  \big]_S 
           \Big]_{JM} \;\Big\} .
\label{eq:h6lwf}
\end{eqnarray}
Here, the operator ${\cal A}$ stands for antisymmetrization
between the two valence neutrons.
In Eq.~(\ref{eq:h6lwf}), the isospin coupling is omitted since $^6_\Lambda$H is streched in isospin space. 
$\chi_{\frac{1}{2}}(n),\chi_{\frac{1}{2}}(t),
\chi_{\frac{1}{2}}(\Lambda)$,
$\eta_{\frac{1}{2}}(n)$ and $\eta_{\frac{1}{2}}(t)$
 are the spin and isospin functions of the
neutron, triton and $\Lambda$, respectively.
It is to be noted that, in Eq.~(\ref{eq:h6lwf}),
the way of spin and isospin coupling is 
the same for all the channels ($c=1 \sim 9$)
 simply to perform numerical calculation easily.
But, we have no problem for the present calculation,
since we employ the full space of the spin and isopin variables for each channel.
The isospin- and spin function employed in Eq.~(\ref{eq:h6lwf})
are also successfully applied to the calculation of four-nucleon bound state of
$^4$He\cite{Kamada01}.

Following the Gaussian Expansion Method (GEM)
\cite{Kami88,Kame89,Hiyama03,Hiyama2012ptep},
we take the functional forms of
$\phi_{nlm}({\bf r})$,
$\psi_{NLM}({\bf R})$ and
$\xi^{(c)}_{\nu\lambda\mu} (\mbox{\boldmath $\rho$}_c)$ as
\begin{eqnarray}
      \phi_{nlm}({\bf r})
      &=&
      r^l \, e^{-(r/r_n)^2}
       Y_{lm}({\widehat {\bf r}})  \;  ,
 \nonumber \\
      \psi_{NLM}({\bf R})
      &=&
       R^L \, e^{-(R/R_N)^2}
       Y_{LM}({\widehat {\bf R}})  \;  ,
 \nonumber \\
      \xi_{\nu\lambda\mu}(\mbox{\boldmath $\rho$})
      &=&
       \rho^\lambda \, e^{-(\rho/\rho_\nu)^2}
       Y_{\lambda\mu}({\widehat {\rhovec}})  \; ,
\end{eqnarray}
where the Gaussian range parameters are chosen 
according to geometric progressions:
\begin{eqnarray}
      r_n
      &=&
      r_1 a^{n-1} \qquad \enspace
      (n=1 - n_{\rm max}) \; ,
\nonumber\\
      R_N
      &=&
      R_1 A^{N-1} \quad
     (N \! =1 - N_{\rm max}) \; ,  
\nonumber\\
      \rho_\nu
      &=&
      \rho_1 \alpha^{\nu-1} \qquad
     (\nu \! =1 - \nu_{\rm max}) \; .  
\end{eqnarray}

The angular momentum space of $l, L, \lambda \leq 2$ 
is found to be sufficient to obtain good convergence of the
calculated results.
Application of the GEM to the
three-, four- and five-body calculations of
single- and double-$\Lambda$ hypernuclei have been extensively 
performed and are reviewed in 
Refs.~\cite{Hiyama2012ptep,Hiyama09PPNP,Hiyama2010,Hiyama2012FBS}.

  The eigenenergies $E$  in Eq.~(1) and the
coefficients $C$ in Eq.~(3) are determined by
diagonalizing the Hamiltonian in Eq.~(2)
with the use of the four-body basis functions introduced above.
If the calculated lowest state is obtained below the lowest-lying
$^4_\Lambda{\rm H}+ n+n$ threshold, it is identified as the
the bound ground state of $^6_\Lambda$H.

When the lowest state is not a bound state but a resonant state,
we calculate the energy and width of the resonance employing 
the stabilization (real-scaling) method~\cite{real-scaling}
that is useful in
calculating narrow resonances 
and is tractable even in the case of four-body resonant states. 
(Note that we take the CSM
when calculating  the broad three-body resonance of $^5$H 
as explained below). 
In the stabilization method, we first diagonalize the Hamiltonian
and obtain the eigenenergies in the same manner as in the bound-state
calculation but changing (scaling) 
the Gaussian range parameters~(5), for example,
as $r_n \to \alpha r_n (\alpha \sim 0.5$ to 2).
The position and width of a resonance can be estimated,
using Eq.~(4) of Ref.~\cite{real-scaling},
from the degree of the stabilization of the eigenenergy 
versus the scaling of the Gaussian range parameters~(5). 
A good example of such a calculation is shown in
Ref.~\cite{pentahiyama} for the study of the 
pentaquark resonances under the scattering boundary condition.

\subsection{The $^5$H nucleus}

We assume that  the $^5$H nucleus is represented by
the $tnn$ three-body system, which is a part of
the entire $tnn\Lambda$ four-body system.
The sets of the Jacobi coordinates describing the $^5$H nucleus
are given by the channels  $c=2$ and $c=5$ in Fig.~1 
but omitting the $\Lambda$ particle
and the coordinate $\rho$.  The wave function of $^5$H is
written with the use of Eq.~(\ref{eq:h6lwf}) but employing
the channels $c=2$ and $c=5$ only and omitting
the amplitudes for the $\Lambda$ particle. 
The Hamiltonian of the $tnn$ subsystem, say $H_{tnn}$, is given by
Eq.~(\ref{eq:hamil6}) without the terms for the $\Lambda$ particle.

As mentioned before, the ground state of $^5$H is
a resonant state which was  observed  at
$E_{\rm r}=1.7 \pm 0.3$ MeV with  $\Gamma=1.9 \pm 0.4$ MeV with respect
to the $tnn$ three-body breakup threshold.
In order to perform a study of such a broad, low-lying three-body resonance,
we employ the CSM~\cite{CSM-ref1,CSM-ref2,CSM-ref3,Ho,Moiseyev}.
The CSM and its application to the nuclear physics problems
are extensively reviewed in Ref.~\cite{Aoyama} and 
references therein. 
Very recently, 
various types of 3$\alpha$ resonances
in $^{12}$C were studied using the CSM~\cite{C12-CSM-2012}.

Using the CSM, one can directly obtain the energy $E_r$ and 
the decay width $\Gamma$ of the $tnn$ three-body resonance
by solving the eigenvalue
problem for the complex scaled Schr\"{o}dinger equation 
with a scaling angle $\theta$: 
\begin{equation}
[H_{tnn}(\theta) -E(\theta)] \Psi_{tnn}(\theta)=0,
\end{equation}
where the boundary condition of the three-body outgoing wave is 
automatically satisfied for the resonance, giving
$E=E_r -i\Gamma/2$ that is, in principle, independent of $\theta$.
The complex scaled Hamiltonian $H_{tnn}(\theta)$ 
is obtained by making the radial coordinate transformation
\begin{equation}
   r_c \to r_c \,e^{i \theta}, \qquad  
R_c \to R_c \,e^{i \theta} 
\quad  (c=2 \;\mbox{and}\; 5 \;\mbox{in Fig.~1} )
\end{equation}
in the Hamiltonian $H_{tnn}$ of the $tnn$ system which is
introduced in Sect. IIIb.

A great advantage of the CSM is that the resonance
states are described with an $L^2$-integrable wave function.
Therefore, the resonance wave function $\Psi_{tnn}$ can be
expanded in terms of the basis functions such as Eqs.~(4) and (5).

\section{Interactions}

For  the $NN$ interaction $V_{NN}$,
we take the AV8$'$ potential~\cite{AV8}.
For the $\Lambda N$  interaction, 
$V_{\Lambda N}$, we employ an effective single-channel
interaction simulating the basic features of the 
Nijmegen model NSC97f~\cite{NSC97}, where
the $\Lambda N$-$\Sigma N$ coupling effects are renormalized
into $\Lambda N$-$\Lambda N$ components:
The potential parameters of the $\Lambda N$ interaction
are listed in Table I(a) in Ref.\cite{Hiyama09} with
the three-range Gaussian potentials which are chosen to reproduce
the $\Lambda N$ scattering phase shifts calculated 
from the NSC97f. Then,
their second-range strengths in the even states of spin-independent and
spin-spin terms are
adjusted so that the calculated energies of the 
$0^+$-$1^+$ doublet states of $^4_\Lambda$H in the $NNN\Lambda$
four-body calculation reproduce the observed
energies of the states.
More importantly, 
this $\Lambda-N$ interaction 
was applied to $^6_{\Lambda}$He
within an $\alpha n \Lambda$ three-body model, 
resulting in $\Lambda$ separation energy
 $B_{\Lambda}=4.21$ MeV which reproduces the observed
$B_{\Lambda}=4.18 \pm 0.10$ MeV. 
This same $\Lambda N$ interaction and the present $NN$
 interaction were already 
applied in the calculation of the binding energy of 
$^7_{\Lambda}$He
 within the framework of an $\alpha \Lambda NN$ 
four-body model, resulting in $B_{\Lambda}=5.48$ MeV
(see the result using 'even' in Fig.2 of Ref.\cite{Hiyama09}.)
which is consistent within the error bar with the recent data of $^7_{\Lambda}$He at 
Jlab\cite{Nakamura2013}.
We thus consider that the employed $\Lambda N$ interactions 
in this work are reasonable.

The interaction $V_{t \Lambda}$ is obtained by folding the 
$\Lambda N$ G-matrix interaction derived from 
the Nijmegen model F(NF)~\cite{NDF}
into the density of the triton cluster, its strength 
being adjusted so as to reproduce the experimental value of
$B_\Lambda(^4_\Lambda$H) within the $t \Lambda$ cluster model.
Also, the spin dependence of the $t \Lambda$ interaction is such
that it reproduces the $0^+$-$1^+$ doublet splitting in $^4_{\Lambda}$H.
The potential parameters are listed in Table I(b) in Ref.\cite{Hiyama09}.

As for the $V_{tn}$, we employ a potential 
proposed in Ref.~\cite{Zhukov}; it is of  Gaussian shape
with dependence on the angular momentum  and spin
of the $tn$ system. The $d$- and $f$-wave
components were taken to be the same as the $s$ and $p$ waves, 
respectively (this was noted afterwards in Ref.~\cite{Zhukov-1998}).
Additionally, we assume the same dependence
for the partial waves with $l \geq 4$ (though their effect
must be negligible), and  
describe the resultant $V_{tn}$ potential in the parity-dependent way:
\begin{eqnarray}
V_{tn}  = \sum_{i=1}^3 
&\Big[& \{ V_{0,{\rm even},i}(r) +V_{ss,{\rm even},i}(r)
\,{\bf S_n \cdot S_t} \}\,\frac{1+P_r}{2} \nonumber \\
& + & \{ V_{0,{\rm odd},i}(r) +V_{ss,{\rm odd},i }(r)
\,{\bf S_n \cdot S_t} \}\,\frac{1-P_r}{2} \nonumber \\
&+ & V_{{\rm SO},i}(r)\, \mbox{\boldmath $\ell$}
{\bf \cdot}({\bf  S_n +S_t} ) \,\frac{1-P_r}{2}
\:\Big]\: e^{-\mu_i r^2},
\label{eq:poten}
\end{eqnarray}
where $P_r$ is the space exchange (Majorana) operator. 
${\bf S_n}=$\mbox{\boldmath $\sigma_n$}/2 and 
 ${\bf S_t}$=\mbox{\boldmath $\sigma_t$}/2,
namely, spin operators for the neutron and triton cluster,
respectively.
The potential parameters of $V_{tn}$ are
listed in Table I.
 
\begin{table}[h]
\caption{Parameters of the 
triton-neutron potential  $V_{tn}$ defined by 
 Eq.~(\ref{eq:poten}).
The parameters are the same as in Ref.~\cite{Zhukov}
except that the components for the partial waves 
$l \geq 4$ are omitted there.
Size parameters are in fm$^{-2}$
and strengths are in MeV.
}
\vskip 0.2cm
 \begin{tabular}{cccccc}
 \hline
 \hline
\noalign{\vskip 0.2 true cm} 
$i$ & & $1$  &$2$  &$3$  \\
\noalign{\vskip 0.2 true cm} 
\hline
$\mu_i$ & &$2.1$  &$\;$ $3.725$ $\;$  &3.015  \\
$V_{0,{\rm even},i}$ & & $\;$ $\;$ 205 $\;$ $\;$ &- &- \\
$V_{ss,{\rm even},i}$ & &60 &-  &- \\
$V_{0,{\rm odd},i}$ & &-  &$-2.2$  &$\;\, -13.95$ @\\
$V_{ss,{\rm odd},i}$ & &- &$8.8$  &$-18.6$ $\;$  \\
$V_{{\rm SO},i}$ & &- &- &4.67 \\
\noalign{\vskip 0.2 true cm} 
\hline
 \end{tabular}
\end{table}
%

However, since the observed energy of the ground state of $^5$H
cannot be reproduced by using the two-body $tn$ potential only,
we introduce an effective three-body force,
$V_{tnn}$ in Eq.~(\ref{eq:hamil6}),
 whose definition is the same as in
Ref.~\cite{Diego}:
\begin{eqnarray}
V_{tnn}=V_{3b}\,e^{-(\rho_{\rm H}/b_{3b})^2},
\end{eqnarray}
where the hyperradius $\rho_{\rm H}$ is defined by
\begin{eqnarray}
  (m_n+m_n+m_t) \, \rho^2_{\rm H} \nonumber 
 = m_n r_{nn}^2 +m_t r_{tn}^2 +m_t r_{tn}^2
\end{eqnarray}
with the obvious notation.
We take a range parameter of $b_{3b}=2.6$ fm
and tune the strength parameter $V_{3b}$ so as to
reproduce the energy and width of $^5$H within the error bar
of the experimental data.
The details are discussed in the next section.
 
\section{Results and discussion}

Before studying the structure of the hypernucleus $^6_{\Lambda}$H,
it is essentially important to succeed in explaining the structure 
of the core nucleus $^5$H to which a $\Lambda$ particle is added.
We solved the CSM equation (6) and 
obtained the  $^5$H ground state as a resonant pole with
$J=1/2^+$.
Taking the strength of the three-body potential as
$V_{3b}=-45$ MeV, we have a resonance pole at
$E_{\rm r}=1.60$ MeV and $\Gamma=2.44$ MeV close to the
central value of the observed energy with the error
($E_{\rm r}=1.7 \pm 0.3$ MeV and
$\Gamma=1.9 \pm 0.4$ MeV), which is illustrated
in Fig.~2(a) for $^5$H (upper).

For  $^6_{\Lambda}$H,
due to the $\Lambda N$ spin-spin interaction,
we have $0^+$-$1^+$ spin-doublet states and the 
$0^+$ state is the ground state.
We discuss how the energy of the $0^+$ ground state of $^6_{\Lambda}$H
changes with respect to  the position of the $^5$H resonance
as the strength $V_{3b}$ of the $tnn$ three-body force is increased.

\begin{figure}[htb]
\begin{center}
\epsfig{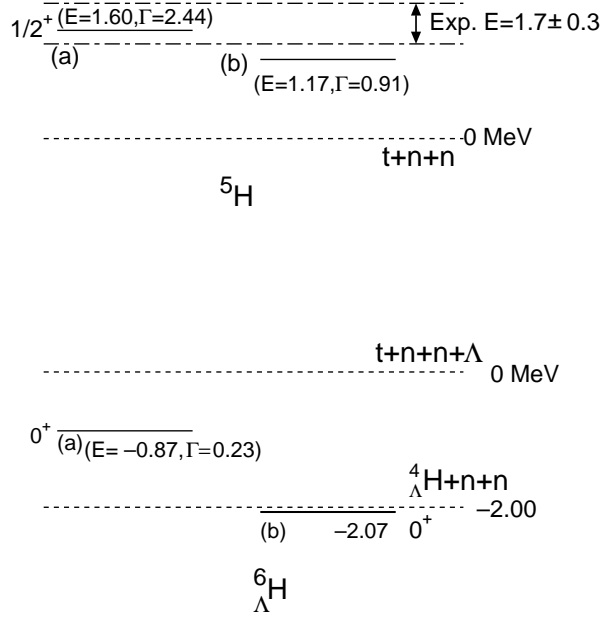}
\end{center}
\caption{Calculated energies and decay widths of $^5$H (upper)
and  $^6_{\Lambda}$H (lower)
in the case that the $tnn$ three-body force~(9)
has a strength  (a) $V_{3b}=-45$ MeV
and (b) $V_{3b}=-65$ MeV.
}
\label{fig:level}
\end{figure}

As shown in Fig. 2(a), 
when the energy of $^5$H is close to the central value
of the observed data, 
the ground state of $^6_{\Lambda}$H is obtained, with $V_{3b}=-45$ MeV, 
at the total energy $E=-0.87$ MeV ($\Gamma=0.23$ MeV), which is 
located by 1.13 MeV above the
$^4_{\Lambda}{\rm H}+n+n$ threshold as a resonance.
In this case, we have  a $\Lambda$ separation energy 
$B_\Lambda=2.47$ MeV with respect to $^5{\rm H}+\Lambda$.

It is expected that the $1^+$ state of $^6_{\Lambda}$H is
located  above the $0^+$ state by about $1$ MeV. 
This is because the two valence neutrons have spin 0 in $^6_{\Lambda}$H,
and the $\Lambda N$ spin-spin
interaction between the two neutrons and the $\Lambda$ particle cancels.
The energy splitting of the $0^+$-$1^+$ doublet states in
$^6_{\Lambda}$H is considered to correspond to that in $^4_{\Lambda}$H.
Therefore, in the case of (a) in Fig.~\ref{fig:level}, 
the $1^+$ state is expected to lie  above
the $tnn\Lambda$ four-body breakup threshold.
At present, it is difficult to precisely calculate the energy and decay 
width of the $1^+$ resonance using the stabilization method. 

As the energy of $^5$H is lowered by increasing the strength
of the three-body force, 
the energy of $^6_{\Lambda}$H becomes closer 
to the lowest  $^4_{\Lambda}{\rm H}+n+n$ threshold. 
However, even if we adjust the calculated energy of $^5$H 
to the lower edge of the error band on the observed energy of $^5$H, 
we cannot  obtain any bound state for $^6_{\Lambda}$H.

When the resonance energy of $^5$H reaches $E_{\rm r}=1.17$ MeV
with $V_{3b}=-65$ MeV, 
we obtain a very weakly bound state of $^6_{\Lambda}$H 
at $E=-2.07$ MeV 
as shown in Fig.~\ref{fig:level}(b).
In this case we have $B_\Lambda=3.24$ MeV with respect to $^5{\rm H}+\Lambda$,
which is consistent with the FINUDA data
$B_{\Lambda}=4.0 \pm 1.1$ MeV~\cite{FINUDA,FINUDA-full}
within the experimental error. 

In order to demonstrate how 
$B_{\Lambda}(^6_\Lambda{\rm H})$ depends 
on the energy and decay width of the core nucleus $^5$H,
we consider one more case.
If we have a $tnn$ three-body force with $V_{3b}=-73$ MeV, 
then we obtain a lower-lying resonance for $^5$H
at $E_{\rm r}=0.73 $ MeV with $\Gamma=0.50$ MeV and 
a bound state of $^6_{\Lambda}$H at 
$E=-3.02$ MeV that is 1.01 MeV below the 
$^4_\Lambda{\rm H}+n+n$ threshold.  
In this case, we have $B_{\Lambda}=3.75$ MeV, which is much 
larger than $B_{\Lambda}=2.47$ MeV
in the case of (a) in Fig.~\ref{fig:level}.
Here, we note that the energy and width of the above artificially
lowered $^5$H resonance is similar to those of the
ground-state resonance of the $^5$He nucleus ($E_{\rm r}=0.80$ MeV with
$\Gamma=0.65$ MeV), but $B_\Lambda$ of $^6_\Lambda$H (3.75 MeV) is
smaller than  $B_\Lambda$ of $^6_\Lambda$He
($4.18 \pm 0.10$ MeV by experiment and 4.21 MeV by our calculation).
 This fact is reasonable. Because,
two valence nucleons in $^5$H are occupied
in the $0p$ orbit.  On the other hand, in $^5$He one is in the 0$p$ orbit
and the other is in $0s$ orbit. Then, this results in the 
weaker $\Lambda N$ attraction (smaller $B_\Lambda$)
 in $^6_\Lambda$H than in $^6_\Lambda$He .

As mentioned before, even if we adjust 
the $tnn$ three-body force so as to match the lower limit 
of the error of the observed energy of $^5$H, 
we could have no bound state for $^6_{\Lambda}$H.
However, by identifying $^6_{\Lambda}$H ground state through 
the observation of its two-body weak-decay $\pi^-$ meson,
 the FINUDA experiment~\cite{FINUDA,FINUDA-full} 
was able to claim that $^6_{\Lambda}$H is a bound system.
Then, in our $tnn\Lambda$ four-body model 
with no explicit $\Lambda N$-$\Sigma N$ coupling effect, the $^5$H resonant 
state should exist above the $tnn$ three-body breakup threshold
by less than $1.17$ MeV. 

Recently, the E-10 experiment to search  $^6_{\Lambda}$H has been
performed at J-PARC and the analysis is now in progress.
If the experiment confirms the existence of this hypernucleus as a bound state,
it is desirable to measure 
the binding energy of the core nucleus $^5$H  
with a precision 
of 100 keV.

Let us compare our results for the $\Lambda$ separation energy 
$B_\Lambda$  with those given in the literature.
The shell model analysis by Dalitz~\cite{Dalitz}
reported  $B_{\Lambda}=4.2$ MeV.
A recent shell model calculation by Millener cited in
Ref.~\cite{FINUDA-full} 
gave  $B_{\Lambda}=4.28 \pm 0.10$ MeV
by using the three binding energies of 
$^5_{\Lambda}$He ($B_{\Lambda}=3.12 \pm 0.02$ MeV),
$^7_{\Lambda}{\rm He}(B_{\Lambda}=5.36 \pm 0.09$ MeV), and
$^4_{\Lambda}{\rm H}(B_{\Lambda}=2.04 \pm 0.04$ MeV).
The results for $B_{\Lambda}$ of $^6_\Lambda$H 
from those shell model calculations are 
similar to the observed data of $^6_{\Lambda}$H.
Akaishi {\it et al}~\cite{Akaishi}, however, obtained
$B_\Lambda=5.8$ MeV, which is by $\sim  1.5$ MeV larger 
than the above two shell-model results.
Each of these three results for $B_\Lambda$ are much larger than the $B_\Lambda$
from the present calculation.
The difference comes from whether or not the core nucleus $^5$H 
is taken into account explicitly in the calculations as a broad three-body resonant state.

Here, it should be reiterated that
$\Lambda N-\Sigma N$
coupling is not explicitly treated in our model.
Since the total isospin of $^5$H is $3/2$, 
it is likely that the coupling 
plays an important role, especially, working as 
an effective $\Lambda NN$ three-body force 
in the binding energies of $^6_{\Lambda}$H.
So far, some authors have investigated the role of
the $\Lambda N-\Sigma N$ coupling in the binding energies
of $^3_{\Lambda}$H, $^4_{\Lambda}$H and $^4_{\Lambda}$He 
\cite{Dalitz58,Gibson72,Gibson88,Miyagawa00,Carlson91,
Akaishi00,Hiyama01,Nemura02}.
For example, in Ref.~\cite{Hiyama01},
four-body calculations for $^4_{\Lambda}$H and $^4_{\Lambda}$He were 
performed taking account of the
$\Lambda N$-$\Sigma N$ coupling explicitly, and it was found that
the effective $\Lambda NN$ three-body force 
contributes about 0.6 MeV to the binding energy of the  $0^+$ ground state
in the case of the NSC97f $YN$ interaction.
As far as the neutron-rich hypernucleus $^6_{\Lambda}$H is concerned, 
Akaishi {\it et al.}~\cite{Akaishi} argued that 
$\Lambda N$-$\Sigma N$ coupling gave a large contribution to 
its binding energy.
On the other hand, a recent shell model calculation by Millener
cited in Ref.~\cite{Millener2012}
found that the $\Lambda N -\Sigma N$ contribution is small in $^6_{\Lambda}$H.
Thus, the determination of the size of the
effect in the binding energy of $^6_{\Lambda}$H is still unsettled.
To answer the question,  
it is necessary to perform  a coupled-channel calculation 
taking into account the $tnn\Lambda$ and $t(^3{\rm He})NN\Sigma$ channels.
This  is one of our future subjects.
The new data from the search for $^6_{\Lambda}$H 
in Experiment E-10 at J-PARC will
provide us with useful information about
$\Lambda N$-$\Sigma N$ coupling.

\begin{figure}[htb]
\begin{center}
\epsfig{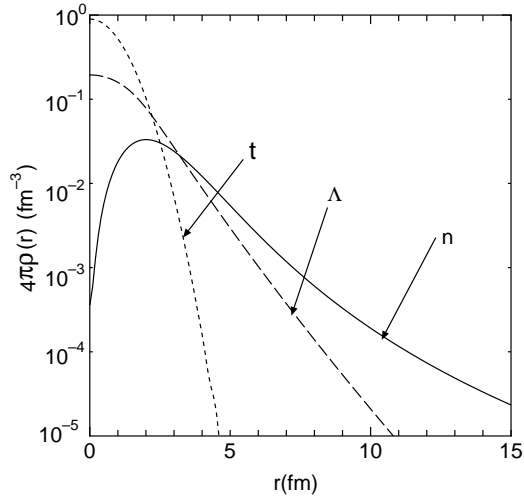}
\end{center}
\caption{Single-particle densities
of the 0$s$ nucleons comprising the triton (dotted curve),
the $\Lambda$ particle (dashed curve) 
and one of the valence neutrons (solid curve) in the ground state of
the hypernucleus $^6_\Lambda$H when the state becomes a bound state
shown in Fig.~2(b).
The radius $r$ is measured from the c.m. of the triton.
}
\label{fig:density}
\end{figure}

Finally, it is interesting to examine the spatial 
structure of the ground state of $^6_{\Lambda}$H 
when the state becomes bound as in
Fig.~2(b). 
Using our wave function of this state, we calculated
the single-particle density distributions of the constituent particles
and plotted them in Fig.~\ref{fig:density};
the dotted curve denotes the density of
the 0$s$ nucleon in the triton, 
the dashed curve is for the $\Lambda$ particle
and the solid line for one of the two valence neutrons.
The $\Lambda$ particle is 
bound to the triton cluster mostly
in the $0s_{\Lambda}$ orbit that is much shallower
than the $0s$~orbit of the nucleon in the triton, 
whereas the valence nucleons are very loosely coupled to the
triton.
Thus, we see, as should be anticipated, that there are three layers in the matter 
distribution of the hypernucleus $^6_{\Lambda}$H,
namely, the nucleons in the well bound triton core, the loosely bound $\Lambda$ 
skin in the $^4_\Lambda$H subsystem, and the two-neutron halo that is unbound 
in $^5$H.

\section{Summary}

We have studied the structure of the 
neutron rich hypernucleus $^6_{\Lambda}$H within
the framework of a $tnn\Lambda$ four-body model.
For this study of $^6_{\Lambda}$H, it is essentially important to reproduce
the property of the ground state of the core nucleus $^5$H which is 
a low-lying, broad three-body resonance at
$E_r=1.7 \pm 0.3$ MeV ($\Gamma=1.9 \pm 0.4$ MeV) with respect to the
$t+n+n$ threshold.
 We note that the $\Lambda$ separation energy, 
$B_\Lambda$, depends strongly on the spatial size of the core 
nucleus~\cite{Hiyama2000PRL,Hiyama2012PTP}.
We thus utilize a $tnn$ three-body model for the $^5$H nucleus and
treat both $^5$H and $^6_\Lambda$H consistently.

In the present $tnn\Lambda$ model, all the two-body interactions
among subunits (triton, two neutrons and $\Lambda$)
are chosen to reproduce the low energy properties, such as
binding energies and scattering phase shifts
of each of the subsystems composed of two and three subunits.
The $NN$ interaction is given by the AV8$'$ potential~\cite{AV8}.
We employ a $\Lambda N$ effective potential
that simulates $\Lambda N$ scattering phase shifts
of the NSC97f interaction~\cite{NSC97} and is slightly tuned to
reproduce the observed energies of the spin-doublet 
$0^+$-$1^+$ states of $^4_{\Lambda}$H.
The $\Lambda N-\Sigma N$
coupling is renormalized into
the $\Lambda N$ potential.
But, we note that there remains an effective $\Lambda NN$ three-body force
that is not renormalizeable into the effective $\Lambda N$ potential.
The employed $\Lambda N$
interaction reproduces the observed binding energies of
$^6_{\Lambda}$He and $^7_{\Lambda}$He\cite{Hiyama09}  within the
framework of $\alpha \Lambda N$ and $\alpha \Lambda NN$ three-
and four-body models, respectively.

The $tn$ potential is taken from Ref.~\cite{Zhukov,Zhukov-1998}.
We found that  
the observed resonance energy of $^5$H cannot be reproduced
with the two-body $NN$ and $tn$ interactions only,
and therefore we introduced a phenomenological, attractive 
$tnn$ three-body force. 
When the  $tnn$ force is tuned 
to reproduce the central value of the observed $^5$H resonance
energy, we obtain $E_r=1.60$ MeV ($\Gamma=2.44$ MeV) and, 
at the same time, we have
the $^6_\Lambda$H ground state  as a resonance
at $E=-0.87$ MeV ($\Gamma=0.23$ MeV) with respect to the
$t+n+n+\Lambda$ threshold; it is located 1.13 MeV above 
the lowest $^4_\Lambda{\rm H}+n+n$ threshold (Fig.~2(a)).
In this case, we have $B_\Lambda=2.47$ MeV. 
Even if the $tnn$ three-body force were adjusted
to reproduce the lower band ($E_{\rm r}=1.4$ MeV) based on the error of
the observed  $^5$H resonance energy, 
we could not obtain any bound state for $^6_{\Lambda}$H
below the $^4_\Lambda{\rm H}+n+n$ threshold.

We were able to generate a shallow bound state  of $^6_{\Lambda}$H
at $E=-2.07$ MeV
(by 0.07 MeV below the $^4_\Lambda{\rm H}+n+n$ threshold)
if the $tnn$ three-body force is tuned to have
the $^5$H resonance at $E=1.17$ MeV ($\Gamma=0.91 $ MeV) 
which is, however, 0.23 MeV below the quoted lower band of 
the error on the observed energy (Fig.~2(b)).
In this case, we have $B_\Lambda=3.24$ MeV, which
is consistent with the FINUDA data ($B_\Lambda=4.0 \pm 1.1$ MeV)
within the error. 

In order to study the structure of $^6_{\Lambda}$H more fully, we
anticipate having a more precise value for the resonance energy of $^5$H 
(100 keV accuracy).
It should be noted that we did not explicitly include 
$\Lambda N-\Sigma N$ coupling in the present work.
If, in the analysis of the E-10 experiment at J-PARC, now in progress,
$^6_{\Lambda}$H is confirmed to exist as a bound state,
we could obtain useful information about $\Lambda N-\Sigma N$ coupling.
To study the effect of such coupling,
it is essential to perform a coupled $tnn\Lambda + t(^3{\rm He})NN\Sigma$
four-body calculation.
This calculation will be one of our future endeavors.

\section*{Acknowledgments}
The authors thank Professors
T. Bressani, S. Marcello, E. Botta, A. Feliciello,
A. Gal and B.F. Gibson for informative discussions.
Also the authors thank Dr. S. Gojuki for support of 
numerical calculation.
The numerical calculations were performed on the 
HITACHI SR16000 at KEK and YITP.
This work was supported by Grants-in-Aid
for Scientific Research from
Monbukagakusho of Japan (No.23224006).

\end{document}